\documentclass{revtex4}

\usepackage{graphicx}
\begin{document}
\newcommand{\nn}{\nonumber}
\newcommand{\be}{\begin{equation}}
\newcommand{\ee}{\end{equation}}
\newcommand{\ba}{\begin{eqnarray}}
\newcommand{\ea}{\end{eqnarray}}
\newcommand{\bann}{\begin{eqnarray*}}
\newcommand{\eann}{\end{eqnarray*}}
\newcommand{\bc}{\begin{center}}
\newcommand{\ec}{\end{center}}
\newcommand{\LEP}{{\sc Lep }}
\newcommand{\LHC}{{\sc Lhc }}
\newcommand{\TESLA}{{\sc Tesla }}
\newcommand{\sts}{\scriptstyle}
\newcommand{\ngs}{\!\!\!\!\!\!}
\newcommand{\rb}[2]{\raisebox{#1}[-#1]{#2}}
\newcommand{\CP}{${\cal CP}$~}
\newcommand{\fgref}[1]{fig.~\ref{#1}}

\newcommand{\sw}{s_{\mbox{\tiny{W}}}}
\newcommand{\cw}{c_{\mbox{\tiny{W}}}}
\newcommand{\MW}{M_{\mathrm{W}}}
\newcommand{\MZ}{M_{\mathrm{Z}}}
\newcommand{\ts}[2]{#1_{\mathrm{#2}}}
\newcommand{\gtot}{g_{\mathrm{tot}}}
\newcommand{\mt}{\tilde{M}}
\newcommand{\smu}{\tilde{\mu}}
\newcommand{\schi}{\tilde{\chi}}

\newcommand{\dmij}{\Delta m^2_{ij}}

\newcommand{\lesim}{\raisebox{-.3ex}{$_{\textstyle <}\atop^{\textstyle\sim}$}}
\newcommand{\gesim}{\raisebox{-.3ex}{$_{\textstyle >}\atop^{\textstyle\sim}$}}
\newcommand{\nslash}{\not{\!n}}
\newcommand{\slsh}[1]{/ \!\!\!\! #1}
\newcommand{\er}{e_{\mathrm{R}}}
\newcommand{\el}{e_{\mathrm{L}}}
\newcommand{\ter}{\tilde{e}_{\mathrm{R}}}
\newcommand{\tel}{\tilde{e}_{\mathrm{L}}}

\bibliographystyle{revtex}

\begin{flushright}
DESY 01-200 \\
Snowmass E3061
\end{flushright}

\title{Slepton Pair Production at a Linear Collider}



\author{A.~Freitas and D.~J.~Miller}
\affiliation{Deutches Elektronen-Synchrotron DESY, D-22603 Hamburg, Germany}


\date{\today}

\begin{abstract}
\noindent Accurate theoretical calculations of slepton pair
production processes at threshold are necessary for an accurate
determination of slepton masses. We discuss the gauge invariant
calculation of these processes for selectron and smuon pairs,
including finite width effects and Coulomb corrections. Energy cuts
to reduce irreducible supersymmetric backgrounds are presented.
\end{abstract}

\maketitle


In any investigation of supersymmetry (SUSY) it is crucial to determine
the masses of the particles to a very high accuracy, since a precise
knowledge of the SUSY spectrum can constrain the various models of
supersymmetry breaking.  In this contribution we will discuss the
measurement of the selectron and smuon masses at a future linear
collider, where the slepton mass can be directly measured from a
threshold scan of slepton pair production. This contribution is based
on the research presented in Ref.\cite{fmz}.\\[0.05cm]

Let us first consider selectron pair production.  Each selectron has
two scalar partners, $\tilde{e}_L$ and $\tilde{e}_R$, corresponding to
the two chiral electron states.  Neglecting mixing effects in the
first and second generations, we regard these chiral states as the
mass eigenstates. In our analysis, we adopt the parameters given by
the SUSY reference point {\em RR2}~\cite{RR2}, for which right-chiral
selectrons decay predominantly into an electron and neutralino, while
the heavier left-handed selectrons may also decay into a neutrino and
chargino if kinematically allowed.

In electron-positron collisions, selectron pairs can be formed by
s-channel $\gamma$ and $Z$ exchange or by t-channel $\tilde{\chi}_i^0$
exchange. The s-channel process, and t-channel process where both
selectrons are of the same chirality, produce selectrons in a P-wave
thereby leading to a cross-section which increases $\sim \beta^3$ near
threshold (where $\beta$ is the selectron velocity in the
centre-of-mass frame). This slow rise makes the threshold cross
section measurement difficult.  S-wave production, where the
cross-section rises with $\beta$, is only possible for mixed chirality
selectron production (i.e. $\ter^+ \tel^-$ or $\tel^+ \ter^-$), which
is clearly less desirable due to the dependence on both selectron
masses. Furthermore, the resulting signals are unfortunately plagued
by many Standard Model (SM) and SUSY background processes, in
particular the production of neutralino and chargino pairs (with their
subsequent cascade decays), rendering them unattractive for the
selectron mass measurement.

In contrast, in $e^-e^-$ collisions, selectron pairs of the same
chirality are produced via t-channel $\tilde{\chi}_i^0$ exchange, as
an S-wave. Consequently the production cross-section near threshold
grows $\sim \beta$, giving a much more easily measured signal.
Furthermore, the $e^-e^-$ initial state switches off the majority of
the contributing background processes. As a result, it is much easier
to measure the selectron mass in $e^-e^-$ collisions than in $e^+e^-$
collisions.

For right-chiral selectrons our signal process is then $e^-e^- \to
\ter^- \ter^- \to e^-e^-\schi_1^0\schi_1^0$ resulting in a signature
of two electrons and missing energy. For left-chiral smuons we have
$e^-e^- \to \tel^- \tel^- \to \nu_e\nu_e\schi_1^-\schi_1^-$ and choose
a leptonic decay for the first chargino, $\schi_1^-\to l^-\bar \nu_l
\schi_1^0$ ($l \neq e$), and a hadronic decay for the other,
$\schi_1^-\to q \bar q^{\prime}\schi_1^0$, resulting in a signature of
$l q \bar q^{\prime}$ and missing energy.
     
In order that the selectron mass may be as accurately measured as
possible, it is essential to have very good accuracy in the
theoretical calculation. To this end, we must include finite selectron
widths and Coulomb rescattering effects. The finite widths are
incorporated by introducing a complex selectron mass, \be
m_{\tilde{e}}^2 \;\to\; M_{\tilde{e}}^2 = m_{\tilde{e}}^2 -
im_{\tilde{e}}\Gamma_{\tilde{e}}, \ee with fixed width
$\Gamma_{\tilde{e}}$. The Coulomb interaction due to photon exchange
between the slowly selectrons also gives large corrections of the
threshold cross-section. At leading order, the correction to the
cross-section for {\em stable particles} is given by the universal
Sommerfeld rescattering correction~\cite{somm}, $\sigma_{\rm
  Born} \to (\alpha\pi / 2 \beta) \, \sigma_{\rm Born}$.  For {\em
  unstable} sleptons produced in a state of angular momentum $l$, this
is modified, at leading order, to,
\be \sigma_{\rm Born} \to \sigma_{\rm Born} \frac{\alpha\pi}{2\beta}
\left[ 1- \frac{2}{\pi} \arctan \frac{|\beta_M|^2 - \beta^2}{2 \beta
    \; \Im m \, \beta_M} \right] \Re e \, \left\{ \left[ \frac{\beta^2 +
    \beta_M^2}{2 \beta^2} \right]^l \right\} \ee
with the generalized velocities $\beta = \frac{1}{s} \sqrt{(s-m_+^2-m_-^2)^2-4 m_+^2 m_-^2}$ and $\beta_M = \sqrt{1- 4 M^2/s}$,
for the (complex) slepton pole mass $M$ and the smuon virtualities
$m_+$ and $m_-$. Initial state radiation from emission of collinear
and soft photons, and beamstrahlung effects were also included. For
a final theoretical prediction which will be used to extract the
slepton masses it is also important to include radiative corrections
to the slepton production cross-sections. While not included in this
analysis, work on these corrections is ongoing~\cite{ayres}.

Due to the complexity of the process and the large number of Feynman
graphs which must be calculated, the computer algebra package
\emph{FeynArts}~\cite{feynarts} was used to calculate helicity
amplitudes. These amplitudes were numerically integrated over the
allowed phase-space using a multi-channel Monte-Carlo approach with
appropriate phase-space mappings. The Monte-Carlo error was reduced by
adaptive weight optimization.

\begin{figure}[tbp]
\includegraphics[width=.48\textwidth,height=.35\textwidth]{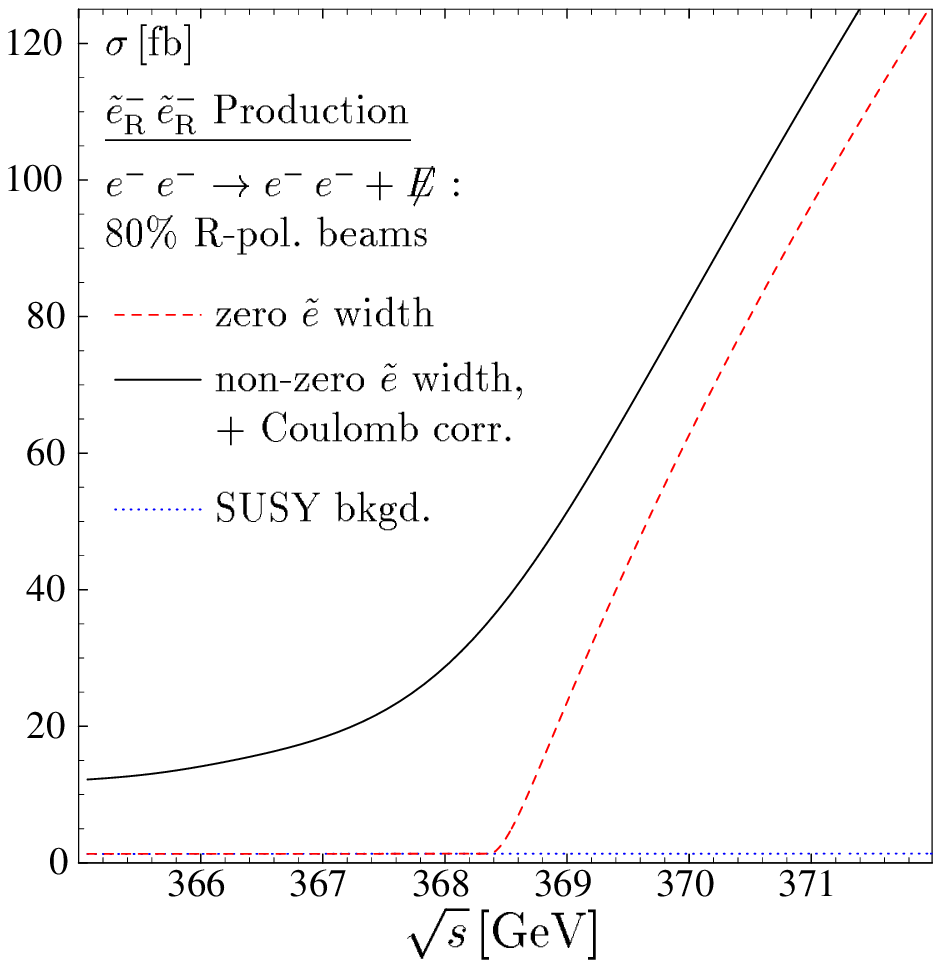}
\includegraphics[width=.48\textwidth,height=.35\textwidth]{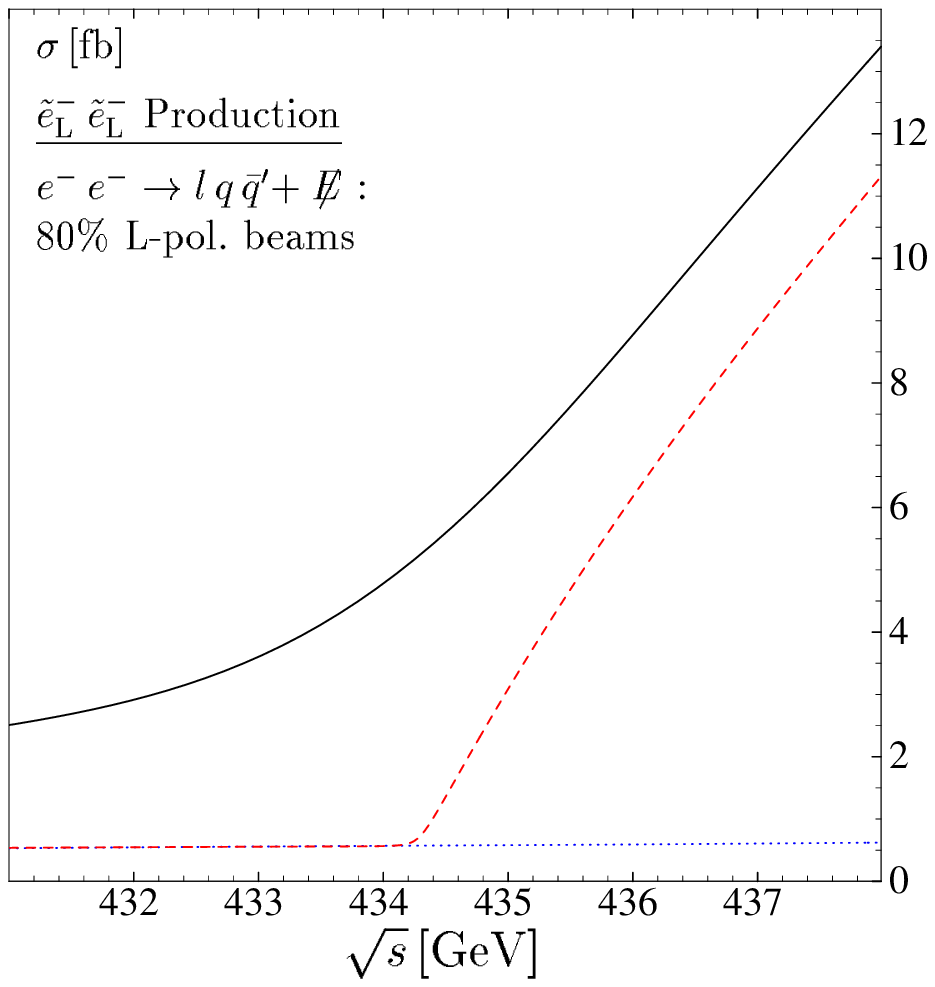}
\caption{\it The signal cross-section and SUSY backgrounds to 
$e^-e^- \to e^-e^- + \slsh{E}$ near the $\ter\ter$ and $\tel\tel$ thresholds. \vspace*{-0.4cm}}
\label{E3061_fig1}
\end{figure}

The threshold cross-sections for right- and left-chiral selectrons in
$e^-e^-$ collisions can be seen in Fig.~(\ref{E3061_fig1}). It is
clear that finite width effects and Coulomb rescattering play an
important role in the selectron mass measurement. Also included are
the important SUSY backgrounds to these processes, which are seen to
be small and very flat over the threshold region, and may therefore be
subtracted in a model independent way by extrapolating from
pre-threshold. Consequently, one can expect that the measurement of
the selectron pair threshold cross-sections in $e^-e^-$ collisions
will provide a very good determination of the selectron masses. For
other recent simulation studies on selectron pair production see
Ref.\cite{fpbm}.\\[0.05cm]

Turning our attention to smuon pair production we find a more
difficult situation. Here there is no t-channel
process; smuon pair production is mediated by s-channel $\gamma/Z$
exchange. The smuons are produced in a P-wave so the cross-section
rises $\sim \beta^3$ at threshold, making its measurement more
challenging than that of the selectron production discussed above.
For this discussion we restrict ourselves to the production of
right-chiral smuon pairs, which are generally of lower mass than their
left-chiral counterparts. For our parameter choice, the dominant decay
is $\smu_R \to \mu \tilde{\chi}_1^0$, giving a signature of two muons
and missing energy.

A problem immediately arises: the doubly resonant process $e^+e^- \to
\smu^+ \smu^- \to \mu^+ \mu^- \tilde{\chi}_1^0 \tilde{\chi}_1^0$ is
gauge dependent for smuons with finite width. Indeed one must include
singly resonant contributions, \mbox{$e^+e^- \to \mu^{\pm} \mu^{\mp \,
    *} \to \mu^{\pm} \smu^{\mp} \schi_1^0 \to \mu^+ \mu^-
  \tilde{\chi}_1^0 \tilde{\chi}_1^0$} and \mbox{$e^+e^- \to
  \tilde{\chi}_1^0 \tilde{\chi}_1^{0 \, *} \to \tilde{\chi}_1^0
  \smu^{\pm} \mu^{\mp} \to \mu^+ \mu^- \tilde{\chi}_1^0
  \tilde{\chi}_1^0$}, to restore gauge invariance. It is this extended
set of Feynman diagrams which we must consider as the signal from
which the smuon mass is to be extracted. For most practical purposes,
this is not a serious concern; one can find 'good' gauges (for example
the Coulomb gauge) where the effect of these extra diagrams is small.
However, for the theoretical precision required at a linear collider,
these extra contributions cannot be discarded. (Notice that this gauge
invariance problem is not manifest in the production of selectrons in
$e^-e^-$ collisions due to its t-channel nature, but would also be
pertinent to s-channel selectron production in $e^+e^-$
collisions.)

We calculate the many Feynman diagrams of signal and background and
perform their integration over phase space using the same methods as
described for the selectrons above. Standard Model backgrounds have
been examined elsewhere and can be removed with appropriate kinematic
cuts~\cite{martyn}. The most important SM background is W boson pair
production, $e^+e^- \to W^+W^-$, where the W bosons decay via $W \to
\mu \nu$. This can be removed by observing that the W decay leptons
lie approximately in an azimuthal plane. The SM process $e^+e^- \to
(\gamma/Z)(\gamma/Z)$ where one $(\gamma/Z)$ decays to muons and the
other to neutrinos, also contributes a sizable background. However,
this can be easily reduced by removing muon pairs which are collinear
(removing $\gamma \to \mu^+\mu^-$) or have an invariant mass
reconstructing the Z (removing $Z \to \mu^+\mu^-$).  These cuts remove
approximately half of the signal.

\begin{figure}[tbp]
\includegraphics[width=.48\textwidth,height=.35\textwidth]{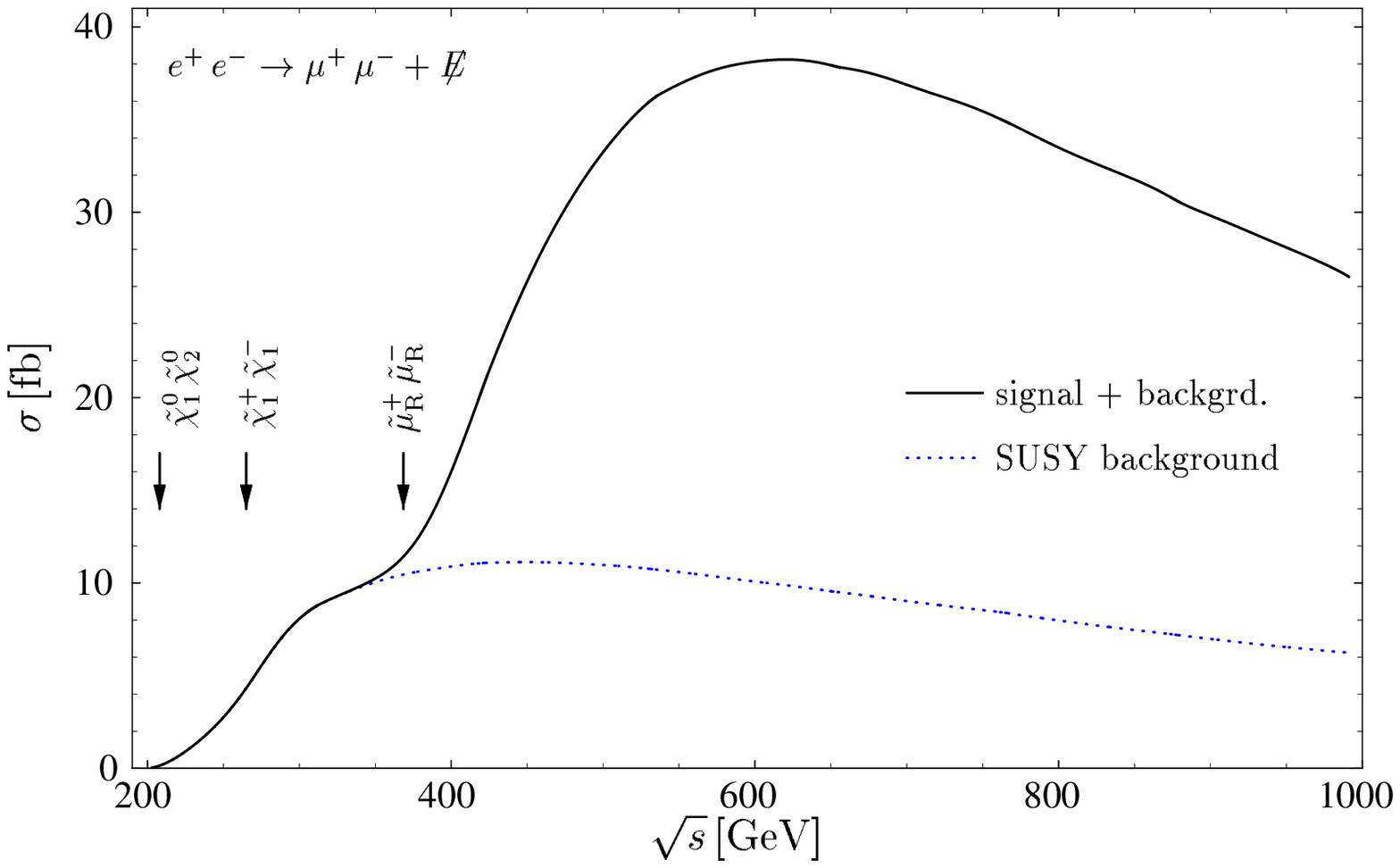}
\includegraphics[width=.48\textwidth,height=.35\textwidth]{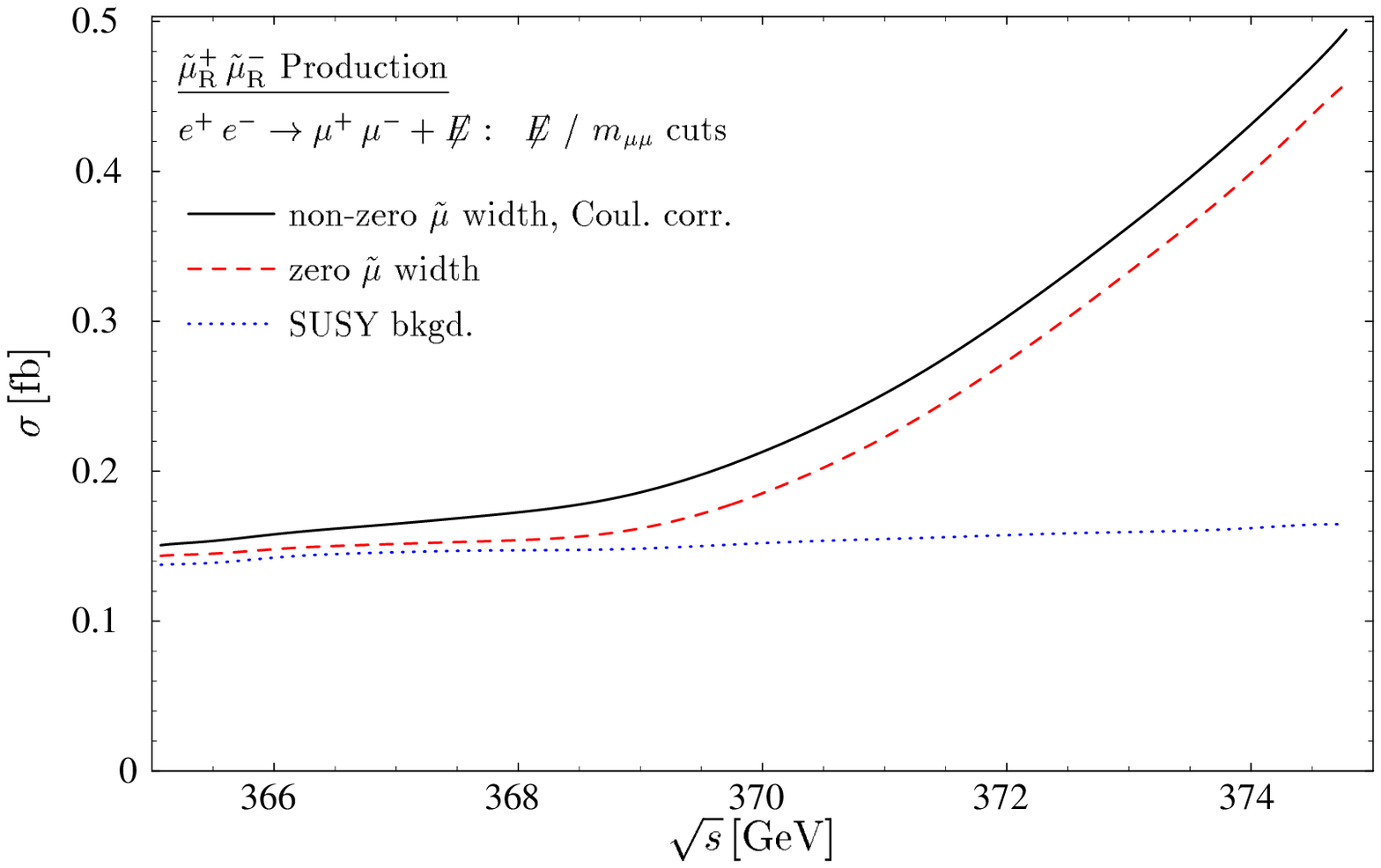}
\caption{(left) The total cross-section and SUSY backgrounds to 
  $e^+e^- \to \mu^+ \mu^- + \slsh{E}$, indicating the important
  nominal thresholds by arrows. (right) \it The cross-sections for signal and associated SUSY backgrounds near the threshold, after the missing energy and muon pair invariant mass cuts. \vspace*{-0.4cm}}
\label{E3061_fig2}
\end{figure}

The MSSM backgrounds are more problematic, since they are large and
appear very similar to the signal. First of all, there is neutralino
pair production, both $e^+e^- \to \schi_k^0 \schi_1^0$ where
the heavier neutralino decays to the lightest one via $\schi_k^0 \to
\mu^+ \mu^- \schi_1^0$, and $e^+e^- \to \schi_2^0 \schi_2^0$ where one
neutralino decays to $\mu^+\mu^- \schi_1^0$ and the other to $\nu \bar
\nu \schi_1^0$. Chargino pair production, $e^+e^- \to \schi_1^+
\schi_1^-$ with the decay $\schi_1^{\pm} \to \smu^{\pm} \nu \schi_1^0$
is also important. Z pair production can also present a SUSY
background if the second Z decays to the invisible neutralinos.
Finally we have Higgs-strahlung, $e^+e^- \to Zh$ where the Z decays to
muons and the Higgs decays to neutralinos.  The contribution of these
backgrounds together with the signal is shown in
Fig.~(\ref{E3061_fig2}/left). The important thresholds are indicted by
arrows.

These SUSY backgrounds can be removed by cuts on the final state's
kinematic properties in two ways. Firstly, the cascade decays will
lead to increased missing energy in the background processes compared
with the signal. This can be optimally exploited by removing all
events with missing energy greater than $0.63 \sqrt{s}$.
Similarly, the muon pair invariant mass will be greater for the signal
than for the large $\schi_2^0 \schi_1^0$ background, which can be
reduced by removing all events with muon pair invariant masses below
$m_{\schi_2^0}-m_{\schi_1^0}$ (approximately $60$ GeV in the case
considered here). Alternatively, one may observe that the muon
energies in the signal contribution will be clustered around their
nominal threshold $(E_{\smu}^2-m_{\schi_1^0}^2)/2m_{\smu}$. The
background exhibits much lower muon energies, so can be removed by
selecting events with muon energies in a band $\Delta E_{\mu} \approx
10$ GeV about this nominal threshold.  The resulting signal and
background after the missing energy and muon invariant mass cuts have
been applied can be seen in Fig.~(\ref{E3061_fig2}/right). Once again,
the remaining background is reasonably flat over the threshold region
and can be removed by extrapolating from below the threshold. Also
note the importance of non-zero width effects and the Coulomb
corrections, where the resulting shift in the cross-section is comparable to 
the expected experimental accuracy.\\[0.05cm]

In summary, the first theoretical steps toward the precise measurement
of the selectron and smuon masses at a linear collider have been
taken. It is clear that the measurement of the selectron masses from
the selectron pair production threshold scan is considerably easier in
$e^-e^-$ collisions than in $e^+e^-$ collisions. In the former case,
the SUSY backgrounds are found to be very small and easily removed by
extrapolation from below threshold. The smuon mass measurement at a
linear collider must rely on $e^+e^-$ collisions, where there are
large SM and SUSY backgrounds. However, these backgrounds can be
removed by the judicious use of cuts on the final state kinematics,
allowing the accurate determination of the smuon mass.


%
%

%
%



\end{document}